\documentclass[aps,PRL,reprint,amsmath,amssymb,superscriptaddress,floatfix]{revtex4-2}
\usepackage{xcolor}
\usepackage{graphicx}
\usepackage{dcolumn}
\usepackage{bm}
\usepackage[colorlinks, citecolor=blue, linkcolor=blue]{hyperref}

\usepackage{orcidlink}

\newcommand{\CNN}{Centre de Nanosciences et de Nanotechnologies, CNRS, Universit\'e Paris-Saclay, 91120 Palaiseau, France}
\newcommand{\LAF}{Laboratoire Albert Fert, CNRS, Thales, Universit\'e Paris-Saclay, 91767 Palaiseau, France}
\newcommand{\LABSTICC}{LabSTICC, CNRS,Universit\'e de Bretagne Occidentale, 29285 Brest, France.}

\begin{document}

\title{Multiple three-magnon splittings in bismuth yttrium iron garnet nanostructures}

\author{Sali Salama \,\orcidlink{0009-0008-4274-0724}}
\email{sali.salama97@gmail.com}
\affiliation{\CNN}
\affiliation{\LAF}
\author{Joo-Von Kim\,\orcidlink{0000-0002-3849-649X}}
\affiliation{\CNN}
\author{Maryam Massouras\,\orcidlink{0000-0002-9518-9285}}
\affiliation{\CNN}
\author{Jamal Ben Youssef\,\orcidlink{0000-0002-9518-9285}}
\affiliation{\LABSTICC}
\author{Abdelmadjid Anane\,\orcidlink{0000-0001-5396-6165}}
\affiliation{\LAF}
\author{Jean-Paul Adam\,\orcidlink{0000-0003-2025-7105}}
\email{jean-paul.adam@universite-paris-saclay.fr}
\affiliation{\CNN}

\date{\today}%

\begin{abstract}
We experimentally demonstrate the generation of multiple three-magnon splitting processes in an in-plane magnetized submicron Bi-YIG disk using micro-focused Brillouin light scattering. The low magnetic damping and strong magneto-optical response of BiYIG enable the detection of nonlinear spin-wave interactions at low threshold powers. By tuning the in-plane static magnetic field, excitation frequency, and power, we observe the generation of three pairs of secondary modes symmetrically distributed around half the excitation frequency. Time-resolved BLS measurements present temporal dynamics and threshold behavior associated with the successive activation of three-magnon pairs.
\end{abstract}

\maketitle

Spin waves and their quanta, magnons, are the elementary excitations of magnetic materials. The inherent nonlinearity of the Landau–Lifshitz–Gilbert equation allows magnons to interact and produce a wide range of nonlinear dynamical phenomena.  This inherent nonlinearity makes spin waves attractive for information processing.  Research has focused on nonlinear effects like multi-magnon scattering and parametric pumping in extended thin films such as yttrium iron garnet (YIG)~\cite{demokritov2006, melkov1996, suhl1957}. More recently, interest has shifted to confined geometries like disks and nano-strips~\cite{demokritov2001, verba_theory_2021, dzyapko2020, guslienko2003dipolar, kim2024stimulated}. In nanostructures, spin waves have quantized frequencies and wave vectors, leading to discrete eigenmodes~\cite{krawczyk2014magnonic}. This spectral selectivity allows precise excitation of individual modes and targeted study of specific spin wave interactions~\cite{hahn2013, srivastava2023identification}. This capacity is essential for advancing magnon-based logic~\cite{chumak2014magnon} and neuroinspired systems \cite{romera2018vowel}.

One promising approach in wave-based computing is reservoir computing.  This involves using physical systems with rich dynamics to process and transform information~\cite{tanaka2019recent}.  For example, nonlinear interactions of spin waves have been shown to be directly applicable to reservoir computing in modal space~\cite{korber2023}.  Here, a set of spin-wave eigenmodes serve as the reservoir’s input states while three-magnon splitting provides the necessary nonlinearity and feedback for computation, resulting in populating a different set of eigenmodes that serve as the output states.  Three-magnon splitting is a nonlinear magnon scattering process where a single magnon at frequency $f$ decays into two lower-energy magnons symmetrically located around $f/2$.  The separation between each mode and $f/2$, denoted by $\delta f$, depends on the dispersion relations and the available magnon modes. 

While many studies focus on individual three-magnon processes~\cite{schultheiss2019whispering}, higher-order nonlinear effects like multiple three-magnon splittings have not been experimentally studied in confined geometries.  The presence of multiple thermal modes around the frequency $f/2$ provides alternative energy conservation pathways, allowing for successive magnon splitting events.  Initially excited at frequency $f$, this mode decays into multiple symmetrically spaced pairs of magnons around $f/2$ with progressively increasing frequency separations: $f/2\pm\delta_1$, $f/2\pm\delta_2$ and $f/2\pm\delta_3$, where $\delta_3>\delta_2>\delta_1$.

This letter investigates nonlinear behavior in submicron disks made of Bi-substituted yttrium iron garnet (BiYIG).  BiYIG was chosen for its exceptionally low magnetic damping and high magneto-optical sensitivity making it ideal for Brillouin light scattering (BLS) studies.  Using micro-focused BLS ($\mu$-BLS), we examine how the spin-wave spectrum changes under various excitation conditions. By adjusting the in-plane bias magnetic field microwave driving frequency and input power we pinpoint the onset of several three-magnon splitting processes.  Time-resolved measurements allow us to track the temporal growth of these modes and determine the thresholds for each pair generated.

\begin{figure}
    \centering
    \includegraphics[width=1
    \linewidth]{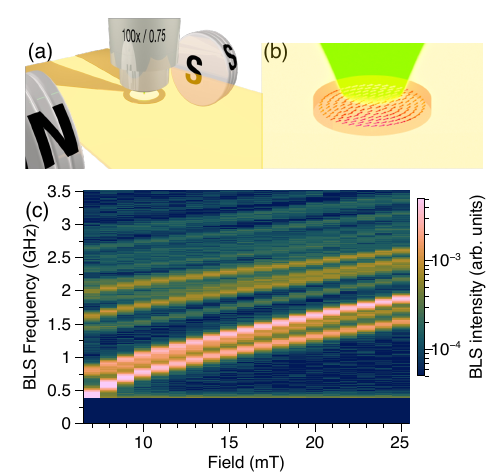}
    \caption{Experimental setup. (a) BiYIG sample is characterized using $\mu$-BLS while applying in-plane static field to get an in-plane magnetized state. (b) The laser spot is focused on the disk center. (c) Thermal BLS spectra as a function of in-plane static magnetic field, varied from 7 mT to 25 mT in steps of 1 mT.}
    \label{fig:art2_setup.pdf}
\end{figure}

Figure~\ref{fig:art2_setup.pdf} illustrates the experimental setup. The sample comprises a 65-nm-thick BiYIG film with a nominal composition of Y$_{2.5}$Bi$_{0.5}$Fe$_{5}$O$_{12}$. This film was grown on a (111)-oriented gadolinium gallium garnet (GGG) substrate using liquid-phase epitaxy. Bismuth was added to enhance the magneto-optical response.  A nanostructured disk with a diameter of 500~nm was defined using electron-beam lithography followed by argon ion-beam etching.  The disk is positioned within an $\Omega$-shaped gold microwave antenna, as depicted in Fig.~\ref{fig:art2_setup.pdf}(a). The dynamic magnetic field, $h_{\mathrm{rf}}$, generated by the radio-frequency current in the antenna is predominantly out of the plane. However, due to a 100-nm lateral offset of the disk from the antenna center, a small in-plane component of $h_{\mathrm{rf}}$ is also present. This field geometry facilitates the excitation of a range of spin-wave symmetries.  A static magnetic field $H_0$ is applied in the plane to achieve an in-plane magnetized state, as shown in Fig.~\ref{fig:art2_setup.pdf}(b). 

$\mu$-BLS is used to detect spin dynamics as illustrated in Fig.~\ref{fig:art2_setup.pdf}(b).  This involves inelastic scattering of photons with magnons. A single mode continuous-wave 532 nm laser is focused at the disk center using a high numerical aperture objective, creating a 300 nm spot diameter.  The backscattered light is analyzed using a tandem Fabry-Pérot interferometer.  $\mu$-BLS allows for the detection of both thermally populated modes and those excited by the radiofrequency (rf) magnetic field.  BLS spectra are recorded in both frequency-resolved and time-resolved modes.  For time-resolved measurements, spin waves are excited at a specific start time $t0$.  The rf excitation is provided as periodic pulses with a fixed duration $\Delta t = 1.18~\mu\text{s}$, controlled by a pulse generator. This generator also simultaneously triggers the time-of-flight analyzer, recording photon arrival times relative to $t_0$.

In the absence of microwave excitation, we first measured the thermally populated spin wave spectrum using $\mu$-BLS as shown in Fig.~\ref{fig:art2_setup.pdf}(c).  All measurements were conducted at room temperature and the spectra were normalized relative to the Rayleigh peak.  The in-plane static magnetic field was varied from 7 to 25 mT in 1 mT increments and the corresponding BLS intensity was recorded at each field step.  Consequently, each spectrum represents the thermally excited eigenmodes at a specific field value. 

The BLS spectra reveal over ten spin wave modes, each exhibiting a distinct frequency increase with the applied in-plane static magnetic field. These modes correspond to quantized standing wave patterns confined within the nanodisk.  The number of nodes along directions parallel and perpendicular to the in-plane field defines these patterns.  The field-dependent frequency shift is consistent with the expected evolution of spin wave dispersion in an in-plane magnetized geometry.

\begin{figure}
    \centering
    \includegraphics[width=1\linewidth]{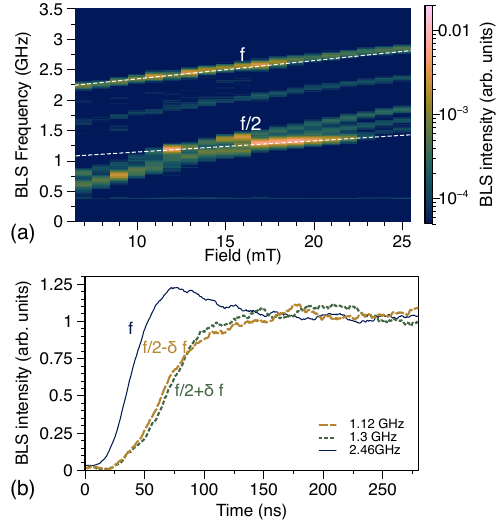}
    \caption{(a)  The BLS spectra were measured as the excitation frequency changed for each field, with the power set at 5 dBm. The mode that was excited is denoted as f. (b) Time-resolved BLS measurements at 14~mT showing the temporal evolution of the spin-wave intensity for the excited mode at 2.46 GHz (solid line) and the secondary modes of three magnon splitting at $f/2-\delta=1.12$ GHz (dashed line) and $f/2+\delta=1.3$ GHz (dotted line). }
    \label{fig:thermal.pdf}
\end{figure}

To investigate the onset of three-magnon splitting, we first identified a mode at H = 14 mT whose frequency $f$ closely matches the sum of two lower-frequency modes around $f/2$.  This mode, labelled $f$ in Fig.~\ref{fig:thermal.pdf}(a), was then tracked in the thermal spectra as the static in-plane field was varied.  At each field value, we excited this mode resonantly using an rf magnetic field. 

Three-magnon splitting is significantly influenced by the applied in-plane static magnetic field. This field shifts the frequencies of spin waves and, due to energy conservation, specific field values enable the generation of modes at or near $f/2$.  At these fields, a magnon at frequency $f$ splits into two lower-frequency magnons near $f/2$, while at other values, this process is suppressed.  Furthermore, three-magnon splitting is governed by symmetry-based selection rules. These rules are determined by the form of the dipolar interaction and the symmetry of the spin wave eigenmodes.

In the field range from 14 to 16 mT, as shown in Fig.~\ref{fig:thermal.pdf}(a), two modes emerge around $f/2$, indicating both energy and symmetry conditions are satisfied for three-magnon splitting.  However, beyond 17 mT and up to 22 mT, only a single mode at $f/2$ is observed.  Further, in the higher field range from 23 mT to 25 mT, no clear nonlinear processes are detected.  Furthermore, the intensity of the directly excited mode decreases for fields above 17 mT suggesting a reduction in coupling efficiency between the rf field and the spin-wave modes. This is consistent with a decrease in the susceptibility with the increase in frequency of the directly-excited mode.

Figure~\ref{fig:thermal.pdf}(b) illustrates a representative example of time-resolved measurement of three-magnon scattering.  At an applied field of 14 mT, the resonant excitation at 2.46 GHz generates two secondary modes at 1.12 and 1.3 GHz.  The temporal evolution of the BLS signal reveals that the directly excited mode is populated first, followed by a delayed rise in the secondary modes. As the excited mode transfers energy to the scattered modes, an overshoot is observed before reaching steady state. This behavior contrasts with parallel and perpendicular parametric pumping, where the amplitude of the directly excited mode remains unaffected by the scattered modes around $f/2$~\cite{merbouche2024degenerate}.

\begin{figure}
    \centering
    \includegraphics[width=1\linewidth]{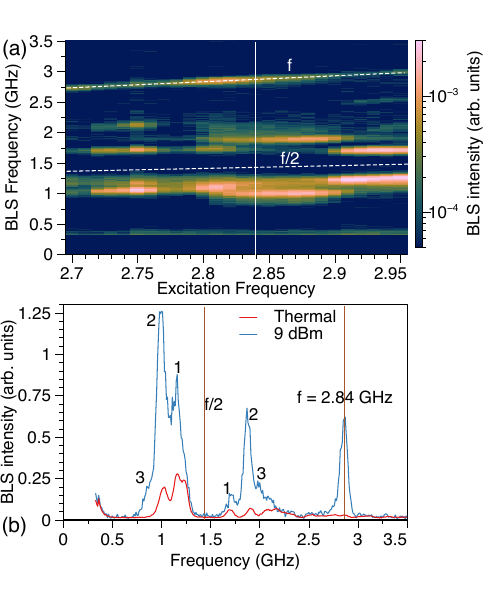}
    \vspace{-0.8cm}
        \caption{BLS spectra measured at an in-plane static magnetic field of 14 mT for excitation frequencies ranging from 2.70 GHz to 2.95 GHz at a constant r.f. power of 9~dBm. (b) BLS spectra at 2.84 GHz and 9~dBm (blue) compared to the thermal level (red).}
    \label{fig:art2_sweepF}
\end{figure}

We focus on the case at 14 mT where two modes emerge around $f/2$. Figure~\ref{fig:art2_sweepF}(a) displays the BLS spectra at an rf power of 9~dBm, measured between 0.4 and 3.5 GHz ($y$-axis) for each excitation frequency between 2.7 and 2.95 GHz in 10 MHz steps ($x$-axis). The color scale represents the BLS intensity. Dashed lines indicate the excitation frequency $f$ and its half $f/2$. This frequency sweep allows us to excite higher-order spin-wave modes and explore their potential to trigger successive nonlinear interactions.

We analyze the spectral evolution and identify regions where multiple secondary modes appear. Between 2.70 and 2.71 GHz, no secondary modes are observed near $f/2$, suggesting that the conditions for three-magnon splitting are not satisfied at this excitation frequency.  From 2.72 to 2.76 GHz, a clear pair of modes emerges at $f/2 \pm \delta_1$. A weaker signal appears at approximately twice the frequency $f/2 - \delta_1$, indicating the presence of a second harmonic.  Between 2.76 and 2.79 GHz, only a single pair of modes near $f/2$ is observed, without higher-order harmonics. However, as the excitation frequency increases to 2.80 to 2.89 GHz, the spectral response becomes richer. Three pairs of modes appear around $f/2$, suggesting the activation of multiple three-magnon splitting channels driven by the same excitation mode. A representative BLS spectrum at 2.84 GHz is shown in Fig.~\ref{fig:art2_sweepF}(b). This spectrum displays three pairs of modes symmetrically distributed around $f/2$. The field and frequency conditions in the 2.80–2.92 GHz range satisfy both the energy matching condition and the spatial symmetry necessary for efficient nonlinear interactions. Finally, from 2.93 to 2.95 GHz, the response simplifies again, with only one pair observed around $f/2$. The appearance of multiple mode pairs within a narrow excitation range indicates the presence of resonant conditions for magnon interactions.

\begin{figure*}
    \centering
    \includegraphics[width=1\linewidth]{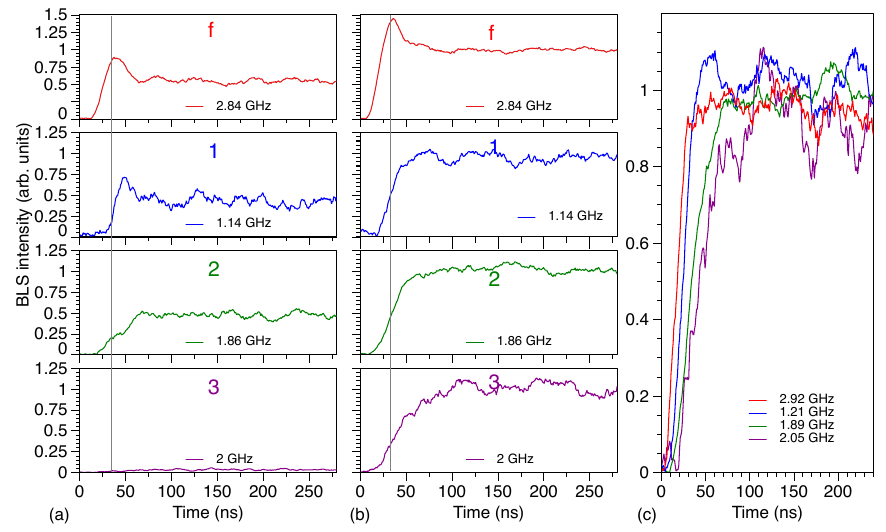}
    \caption{Temporal evolution of the excited mode at 2.84 GHz and the secondary modes generated from three-magnon splitting (labeled “1”, “2”, and “3”) under a static in-plane field of 14 mT at input powers of (a) 6 dBm and (b) 10 dBm. (c) Temporal evolution at an excitation frequency of 2.92 GHz and an input power of 10 dBm, showing a similar sequential population behavior.}
    \label{fig:timeResolved}
\end{figure*}

To understand the three generated magnon pairs, we conducted time-resolved measurements.  For clarity, we plot the time evolution of one representative mode from each pair as shown in Fig.~\ref{fig:timeResolved}.  
Focusing on the time interval between 0 and 260 ns in Fig.~\ref{fig:timeResolved} helps us identify the sequence of mode population. At an excitation frequency of 2.84 GHz and an rf power of 6 dBm [Fig.~\ref{fig:timeResolved}(a)], we observe that the directly excited mode is populated first, as expected. Among the three secondary modes of magnon splitting, the mode closest in frequency to $f/2$, labeled as mode ``1'', appears shortly after the pump onset. Mode ``2'', which is farther from $f/2$, is populated later. The third mode pair, ``3'', remains below the detection threshold at this power level, indicating insufficient energy to excite all three pairs at 6 dBm.

At an increased rf power of 10 dBm [Fig.~\ref{fig:timeResolved}(b)], we observe the excitation of all three secondary mode pairs.  As illustrated in Fig.~\ref{fig:timeResolved}(b), the directly excited mode is populated first. The sequence of appearance for the three secondary modes is: mode 1 (blue), closest to $f/2$, appears first; mode 2 (green) follows shortly after; and finally, mode 3 (purple), the farthest from $f/2$, appears last. A similar sequential population of secondary modes was observed at an excitation frequency of 2.92 GHz. In this case, three mode pairs appear around the frequency $f/2$. As shown in Fig.~\ref{fig:timeResolved}(c), the modes closest to $f/2$ are populated first, followed sequentially by modes located further away in frequency.

The decay of an excited magnon mode at frequency $f$ into secondary modes can be likened to a
magnetic analog of the parametric down-conversion process in nonlinear optics, in which one photon at $f_0$ splits into a pair of photons at $f_{s}, f_{i}$, with $f_s + f_i \approx f_0$~\cite{yariv2007photonics}. In the magnetic system, the pump mode of amplitude $a_0(t)$ couples to several magnon pairs $a_{n-}(t)$ and $a_{n+}(t)$, whose frequencies satisfy the same relation. To model this behavior, we numerically solved a system of coupled-mode equations describing the temporal evolution of the pump and secondary magnon amplitudes~\cite{powers2017fundamentals,yariv2007photonics,Verba2021ThreeMagnon}. The model incorporates the linear damping rates of all modes, their nonlinear saturation terms, the three-magnon coupling strengths between the pump and each generated pair, and intrinsic thermal fluctuations represented by complex Gaussian noise sources. Further details are provided in the Supplementary Material.

Figure~\ref{fig:model_time} presents the simulated intensities of the pump mode and the three secondary mode pairs for two excitation powers. At low power, only the first mode pair at 1.14~GHz and the second pair at 1.86 GHz are generated. As the pump amplitude increases, an additional pair at 2.00~GHz appears and leads to a sequence of three-magnon splittings. The time-resolved profiles show that higher-order pairs appear with a delayed onset compared to the directly excited mode, in agreement with the experimental dynamics observed by $\mu$-BLS. 
\begin{figure}
    \centering
    \includegraphics[width=1\linewidth]{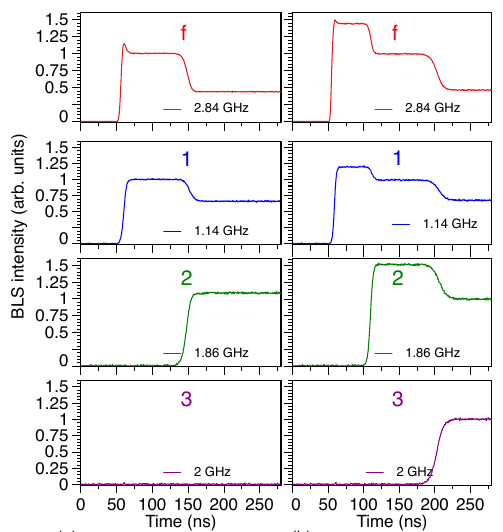}
    \caption{%
    Simulated time-dependent intensities of the pump and secondary modes obtained from the nonlinear three-magnon model. 
    (a) low power excitation, where only the first and second pairs are generated. 
    (b) high power excitation, where three mode pairs appear sequentially around $f/2$. 
    The model reproduces the delayed onset and power-dependent activation of successive magnon pairs observed experimentally.}
    \label{fig:model_time}
\end{figure}
We observe a progressive decrease in the intensity of the pumped mode each time a new secondary pair is generated, which indicates a successive energy transfer from the pump to the generated modes. Similarly, the intensity of pair~1 decreases when pair~2 starts to grow, and the intensity of pair~2 is reduced upon the generation of pair~3, demonstrating the redistribution of energy between the generated magnon pairs.

We have experimentally observed multiple three-magnon splitting processes in a submicron BiYIG disk using $\mu$-BLS.  By adjusting the static magnetic field, rf excitation frequency, and rf power, we observed the sequential generation of multiple magnon mode pairs symmetrically distributed around half the pump frequency. These nonlinear processes occur under specific conditions that satisfy energy conservation and symmetry selection rules.  Time-resolved measurements confirmed that secondary modes appear after the directly excited one and at higher powers, three pairs of modes are observed.  A proposed model based on coupled-mode equations successfully reproduces the observed behavior including the generation of these modes which depends on the threshold and their temporal evolution.

\begin{acknowledgments}
We thank Ping Che and Nathan Beaulieu for valuable discussions and their help with sample preparation. Samples were patterned within the C2N micro and nanotechnologies platforms. We also thank the C2N cleanroom staff for their assistance in sample fabrication. This work was supported by the French RENATECH network and by a public grant overseen by the Agence Nationale de Recherche as part of the ``Investissements d’Avenir'' program (Labex NanoSaclay, contract no. ANR-10-LABX-0035), SPICY, and contract No. ANR-20-CE24-0012 (MARIN). 
\end{acknowledgments}

\bibliography{articles.bib}

\end{document}